\documentclass[10pt, twocolumn, letterpaper, journal]{IEEEtran}
\usepackage{amssymb,amsmath,epsfig,tabularx,delarray,enumerate,graphicx,array,cite}
\usepackage{algorithm}
\usepackage{subcaption}
\usepackage{caption}
\usepackage{algorithmicx}
\usepackage{algpseudocode}
\usepackage{graphicx}
\usepackage{tikz}
\usetikzlibrary{shapes.geometric}
\usetikzlibrary{calc}
\begin{document}
\title{Multi-Hypothesis Compressed Video Sensing Technique}
\author{Masoumeh Azghani, Mostafa Karimi, and Farokh Marvasti$^\ast$
\thanks{$^\ast$M. Azghani, M. Karimi, and F. Marvasti are with Advanced Communications Research Institute (ACRI), Electrical Engineering Department of Sharif University of Technology, Iran (e-mail: azghani@ee.sharif.ir; marvasti@sharif.edu)}}

\markboth{Submitted to IEEE Transactions on Circuits and Systems for Video Technology}{Shell \MakeLowercase{\textit{et al.}}:
Multi-Hypothesis Compressed Video Sensing Technique}

\maketitle
\begin{abstract}

\begin{small}
In this paper, we present a compressive sampling and Multi-Hypothesis (MH) reconstruction strategy for video sequences which has a rather simple encoder, while the decoding system is not that complex. We introduce a convex cost function that incorporates the MH technique with the sparsity constraint and the Tikhonov regularization. Consequently, we derive a new iterative algorithm based on these criteria. This algorithm surpasses its counterparts (Elasticnet and Tikhonov) in the recovery performance. Besides it is computationally much faster than the Elasticnet and comparable to the Tikhonov. Our extensive simulation results confirm these claims. 

\end{small}
\end{abstract}


\section{Introduction}
In conventional video compression techniques, the frames sampled at Nyquist rate are compressed by exploiting their temporal and spatial correlations and applying an entropy coding and quantization. The focus of this paper is Compressed Video Sensing (CVS) which refers to the sampling of the video frames based on the Compressed Sensing (CS) technique.  

Compressed sensing \cite{candes2006compressive, donoho2006compressed, candes2008introduction} has revolutionized the signal sampling and processing systems by integrating the compression and sensing. CS has been applied in various fields \cite{marvasti2009unified} such as image/signal compression, remote sensing, machine learning tasks like clustering \cite{elhamifar2009sparse}, spectrum sensing and channel estimation \cite{pakrooh2012ofdm} in wireless channels, and signal acqusition in wireless sensor networks. The property of the signal utilized by the CS theory is its sparsity which refers to the case where most of the signal coefficients are zero. A large number of the signals we are dealing with can be approximated as sparse in some domain such as (Discrete Fourier Transform) DFT, (Discrete Cosine Transform) DCT, (Discrete Wavelet Transform) DWT, or time domain. As a concrete example, the natural images, speech signals, and video frames are approximately sparse in DCT and DWT domains. This fact encourages us to explore the applicablity of CS in video compression and transmission.\

The application of CS for video, CVS has been studied from different aspects. The main assumption in CVS is that the compressive samples of the frames (a number of linear combinations of the samples) are available at the encoder side and the various CVS schemes attempt to reconstruct the frames from those compressive samples at the decoder side. Moreover, it should be noted here that CS, and consequently CVS, refer to the sampling reduction. The conventional video coding schemes, however, refer to the entropy coding, quantization and data compression.

In \cite{prades2009distributed}, video frames are divided into two groups: Key-frames and CS-frames. The Key-frames are encoded by traditional MPEG, while CS-frames are sensed using a CS measurement matrix. In \cite{kang2009distributed}, Key-frames are reconstructed using GPSR \cite{figueiredo2007gradient}.  The Block-based Compressed Sensing with the aid of Smoothed Projected Landweber reconstruction (BCS-SPL) \cite{mun2009block, mun2011residual} deployed in DWT provides much faster reconstruction than frame-based CS sampling. These methods are known as Single-Hypothesis Motion Compensation (SH-MC) schemes, since they predict each block of the current frame with just one block in the previous reconstructed frames. 
 
 In SH-MC methods, also called block matching, the encoder searches for the block in the reference frame with the highest matching for a specific block in the current frame. The SH-MC has some disadvantages that makes it improper for use. It imposes a transmission overhead on the encoder to send the Block Motion Vectors (BMV) besides the increase in the computational complexity at the encoder-side due to the motion estimation search. Moreover, by estimating each block of the current frame with one in the reference, the SH-MC implicitly assumes that the motions occurring in the video frames are of uniform block translational model. As this assumption does not always hold, the blocking artifacts appear in the recovered frame. To address these issues, the Multi-Hypothesis Motion Compensation (MH-MC) technique has been proposed which transfers the motion estimation task from the encoder to the decoder, eliminating the transmission overhead of the BMVs and simplifying the encoding structure \cite{5749477}. Each block of the current frame is estimated by a linear combination of a number of its surrounding blocks in the reference frame at the decoder end. Thus, the motion compensation accuracy is increased and the blocking artifacts are eliminated. 
 
The MH-MC techniques improve the recovery performance at the expense of more complexity at the decoder side. In \cite{5749477}, the MH-MC approach was considered and due to the ill-posed nature of CVS problem, a simple Tikhonov regularization is proposed. Tikhonov-regularization reconstruction provides higher PSNR as compared to BCS-SPL recovery. In \cite{zhu2013adaptive}, a combination of the MH and SH reconstruction schemes are used. Elasticnet based MH-MC was suggested in \cite{chen2013elastic} which achieves acceptable performance at the expense of more complexity compared to Tikhonov regularizaion reconstruction.\\

 In \cite{5749477}, the sparsity is not exploited to recover the compressive measurements. In \cite{chen2013elastic}, however, the coefficient vector (the vector containing the coefficients of the linear combination of the blocks) is assumed to be sparse, which is not a realistic assumption in general leading to poor recovery performance. Furthermore, the ordinary CS solvers such as Elasticnet have been adopted to recover the frames. Such solvers require the measurement matrix to be under complete. Hence, the authors in \cite{chen2013elastic} have been forced to increase the dimension of the estimating vector as the sampling rate goes up to make the problem under complete. The computational complexity of the solvers increases in an exponential order with the increase of the vector dimension. \\
 
In this paper, we consider the sparsity of the frames which is more realistic than the sparsity of the coefficient vector and combine this with the Tikhonov regularizer. Hence, the recovery quality of the proposed technique is improved. Instead of applying the existing CS solvers, we design a recovery scheme for our model which works even in the case of small-size coefficient vectors. We suggest a new cost function which combines the sparsity condition with the Tikhonov regularizer. The cost function is minimized using the Alternating Direction Method of Multipliers (ADMM) \cite{boyd2011distributed}. The proposed MH-MC scheme performs better than Tikhonov and Elasticnet in terms of Peak Signal to Noise Ratio (PSNR), and is considerably faster than Elasticnet and slightly slower than Tikhonov.

The rest of the paper is organized as follows: 
Section II gives an overview of compressed sensing. The proposed compressed video sensing method is illustarted in section III. The extenssive simulation results are reported in section IV and section V concludes the work. 

\section{Compressed sensing overview}
Let $\mathbf{s}$ be an $n$ dimensional signal sampled by an $m \times n$ meaurement matrix $\mathbf{\Phi}$ ($m<n$) . The measurement vector $\mathbf{y}$ is obtained as:
\begin{equation}\label{ys}
 \mathbf{y}=\mathbf{\Phi}\mathbf{s}
\end{equation}
Compressed Sensing attempts to reconstruct the signal vector $\mathbf{s}$ from an $m$ dimensional set of its measurements $\left(m<n\right)$ using the extra information that the signal is sparse in some domain like $\mathbf{\Psi}$ \cite{candes2006stable}. Sparsity refers to the case where most of the signal coefficients are zero. Assuming that the signal $\mathbf{s}$ is sparse in $\mathbf{\Psi}$ domain, we can have:

\begin{equation}
 \mathbf{s}=\mathbf{\Psi} \mathbf{x}
\end{equation}
where $\mathbf{x}$, indicates the transform coefficients vector of $\mathbf{s}$, say its DCT coefficients. Then, the under-determined set of equations \eqref{ys} can be translated as:
\begin{equation}\label{yx}
 \mathbf{y}=\mathbf{\Phi}\mathbf{\Psi}\mathbf{x}=\mathbf{A}\mathbf{x}
\end{equation}
where $ \mathbf{A}=\mathbf{\Phi}\mathbf{\Psi}$. We solve \eqref{yx} with the sparsity constraint of $\mathbf{x}$. Among all the solutions of \eqref{yx}, CS seeks for the one with most number of zeros. To become more concrete, sparsity is defined as the number of non-zero entries of a signal, \emph{i.e.} the $L_{0}$ semi-norm of the signal. Therefore, CS can be solved using the following:
\begin{equation}\label{P0}
\min_{x}\Vert \mathbf{x} \Vert_{0}   \text{  subject to  }        \mathbf{y}=\mathbf{A}\mathbf{x}
\end{equation}
The only way to solve \eqref{P0} is to conduct an exhaustive search which is $NP$-hard. A tractable approach to solve the non-convex problem \eqref{P0} is to approximate it by its nearest convex problem, which is based on $L_{1}$ norm as:
\begin{equation}\label{P1}
\min_{x}\Vert \mathbf{x} \Vert_{1}          \text{  subject to  }         \mathbf{y}=\mathbf{A}\mathbf{x}
\end{equation}

The problem above is called Basis Pursuit (BP) \cite{chen1998atomic}. LASSO \cite{tibshirani1996regression} is another convex optimization based CS recovery technique which solves lagrangian form of the above problem. 

\section{The Compressed Video Sensing Scheme}
In this section, we would illustrate the proposed scheme for compressed video sensing. The video frames are grouped as reference and non-reference ones for which the compression scheme differs. Moreover, the frames would be encoded and sent to the receiver in block-wise fasion, \emph{i.e.} the non-overlapping blocks of the frames are coded separately. The motion estimation is conducted to remove the temporal or inter-frame redundancy between successive frames. However, as our goal is to simplify the encoding system, the computationally exhaustive motion estimation task is done at the decoder side using the notions of MH motion estimation technique.

\subsection{The searching strategy} \label{serchsec} 
For each block of the non-reference frames, the corresponding $\mathbf{H}$ matrix is constructed by stacking (column-wise) a number of vectorized blocks of the reference frame that are lying inside the search window. For better illustration, a sample of the searching window is depicted in blue in Figure 1. The red square is a block of the reference frame corresponding to a particular block of the non-reference frame. The blue squares indicate the overlapping blocks inside the search window which would be used as estimators of the non-reference block. We select $p$ of those overlapping blocks (in a symmetric manner) and vectorize them to construct the matrix $H$ of size $L^{2} \times p$.

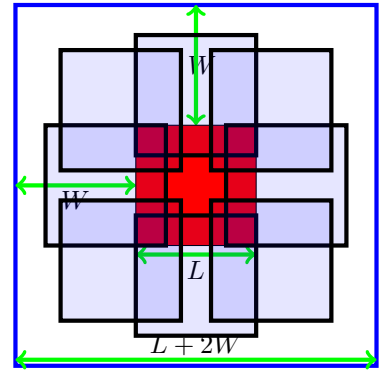
\begin{figure}
\begin{center}
\begin{tikzpicture}[scale=0.4]

\node (x1) at ($(6,3.2cm)$){$L$};
\node (x1) at ($(6,0.7cm)$){$L+2W$};
\node (x1) at ($(6.2,10cm)$){$W$};
\node (x1) at ($(2,5.5cm)$){$W$};
\draw[ultra thick,blue] (0,0) rectangle (12,12);
\draw[ultra thick,green,<->] (4,6) -- (0,6);
\draw[ultra thick,green,<->] (6,8) -- (6,12);
\draw[ultra thick,green,<->] (4,3.7) -- (8,3.7);
\draw[ultra thick,green,<->] (0,0.2) -- (12,0.2);
\draw[fill=red] (4,4) rectangle (8,8);
\draw[fill=blue, ultra thick, fill opacity=0.1] (4,7) rectangle (8,11);
\draw[fill=blue, ultra thick, fill opacity=0.1] (4,1) rectangle (8,5);
\draw[fill=blue, ultra thick, fill opacity=0.1] (1,4) rectangle (5,8);
\draw[fill=blue, ultra thick, fill opacity=0.1] (7,4) rectangle (11,8);
\draw[fill=blue, ultra thick, fill opacity=0.1] (6.5,6.5) rectangle (10.5,10.5);
\draw[fill=blue, ultra thick, fill opacity=0.1] (1.5,1.5) rectangle (5.5,5.5);
\draw[fill=blue, ultra thick, fill opacity=0.1] (1.5,6.5) rectangle (5.5,10.5);
\draw[fill=blue, ultra thick, fill opacity=0.1] (6.5,1.5) rectangle (10.5,5.5);
\end{tikzpicture}
\end{center}\label{pic}
\caption{The schematic of a search window}
\end{figure}
For the simulations, we consider $L=16$ and two selections of $p$ as $p=20$ and $p=400$.

\subsection{The reference frame transmision}
The reference frames are encoded using block compressive sensing. The frame is divided into non-overlapping blocks of size $L\times L$. The blocks are then vectorized row-wise and compressively sensed using the measurement matrix, $\mathbf{\Phi}$, and sent to the decoder. In the receiver, the fundamental signal is recovered using the BCS-SPL recovery algorithm. The recovered vectors are then reshaped to construct the reference frame. The transmission of non-reference video frames is illustrated in the next subsection.

\subsection{The non-reference frame transmission}
The encoder takes compressive measurements of the vectorized blocks of size $L\times L$ and sends them to the decoder where the proposed sparse MH technique is used to recover the blocks. Let's assume that the vector $\mathbf{x}$ of size $n\times 1$ represents the vectorized form of the current block  $\left(n=L^{2}\right)$. The compressive measurements can be obtained as:
\begin{equation}
\mathbf{y}=\mathbf{\Phi}{\mathbf{x}}\label{apb}
\end{equation}
where $\mathbf{\Phi}$ is the $m\times n$ random Gaussian measurement matrix. Let's assume that each current block can be represented as a linear combination of $p$ blocks of the reference frame lying inside the corresponding search window. Thus, the vectorized current block can be as:
\begin{equation}
{\mathbf{x}}=\mathbf{H}{\mathbf{\omega}}
\end{equation}
where $H$ is its corresponding search matrix of size $n \times p$ defined in subsection \ref{serchsec} and $\mathbf{\omega}$ is the coefficient vector. Hence, the measurement vector, $\mathbf{y}$ can be written as:
\begin{equation}
\mathbf{y}=\mathbf{\Phi}\mathbf{H}{\mathbf{\omega}}\label{apblock}
\end{equation}

At the decoder side, the goal is to recover the current block. Knowing that the current block is sparse in the DCT domain ($\mathbf{\Psi}^{-1}$ domain). The $\mathbf{\Psi}$ matrix respresents the inverse DCT matrix. We can recover the vectorized block, ${\mathbf{x}}$ using notions of sparsity. In order to reconstruct the coefficient vector $\mathbf{\omega}$ which leads to the recovery of the current block, we propose to solve the following optimization problem as:

\begin{equation}
\tilde{\mathbf{\omega}}=\text{argmin}_{\mathbf{\omega}} \Vert \mathbf{y}-\mathbf{A}\mathbf{\omega}\Vert_{2}^{2}+\lambda_{1}\Vert \mathbf{\Gamma}\mathbf{\omega}\Vert_{2}^{2}+\lambda_{2}\Vert \mathbf{B}\mathbf{\omega}\Vert_{0}\label{op1}
\end{equation}

where $\mathbf{B}=\mathbf{\Psi}^{-1}\mathbf{H}$ represents the domain in which $\mathbf{\omega}$ is sparse, and $\mathbf{A}=\mathbf{\Phi}\mathbf{H}$.
The first term of the cost function in \eqref{op1} indicates the error. The second term is the Tikhonov regularization term and the matrix $\mathbf{\Gamma}$ is defined as:
\begin{equation}
\mathbf{\Gamma}=diag\left(\Vert\mathbf{y}-\mathbf{\Phi}\mathbf{h}_{1}\Vert_{2}, \cdots,\Vert\mathbf{y}-\mathbf{\Phi}\mathbf{h}_{p}\Vert_{2}\right) 
\end{equation}
where the vectors $\mathbf{h}_{1}, \cdots, \mathbf{h}_{p}$ are the columns of the matrix $\mathbf{H}$.

The last term in \eqref{op1} promotes the sparsity of the current block in the DCT domain. The block is then reconstructed at the decoder side as:

\begin{equation}
\tilde{\mathbf{x}}=\mathbf{H}{\tilde{\mathbf{\omega}}}
\end{equation}

As the $L_{0}$ norm is non-convex, we approximate it with its best convex relaxation, $L_{1}$ norm, thus the ultimate cost function is convex as in \eqref{costadm} which should be solved.
\begin{equation}
\Vert \mathbf{y}-\mathbf{A}\mathbf{\omega}\Vert_{2}^{2}+\lambda_{1}\Vert \mathbf{\Gamma}\mathbf{\omega}\Vert_{2}^{2}+\lambda_{2}\Vert \mathbf{B}\mathbf{\omega}\Vert_{1}
\label{costadm}
\end{equation}
We have considered the sparsity of the linear combination of the reference blocks in the transform domain (such as DCT). It should be noted that our proposed cost function in \eqref{op1} assumes the reconstructed signal to be sparse in the DCT domain and $\mathbf{\omega}$ to be sparse in $\mathbf{B}$ domain which is more realistic than the sparsity assumption of $\mathbf{\omega}$ made in \cite{chen2013elastic}.
The size of the $\mathbf{B}$ matrix is $n \times p$ where $n$ may be greater than $p$, for example in the scenario of $p=20, n=64$ is considered in \ref{sim2sec}. Thus, the dictionary $\mathbf{B}$ may be overcomplete for which we cannot use the ordinary CS recovery techniques such as LASSO. This is because in order to implement such techniques, we should iterate between the sparsity domain and the signal domain which are not equidimensional. Transfering from a lower dimensional space to a higher dimensional one is not a big deal, however, the main problem occurs when transfering back from the lower dimension to the higher one since it would not be a one to one map. To solve this problem, we take advantage of the ADMM technique \cite{han2012note, hong2012linear} which decomposes the cost function into two sub-functions and minimizes each of them with respect to one of the variables. This way, each function is minimized with respect to its own variable and no transfering from one space to the other would be required. 
The ADMM technique parses the cost function into two parts by introducing an auxiliary variable. Then, an alternating iterative scheme is adopted to minimize the cost function. The auxiliary variable in this problem is $\mathbf{x}$ and the cost function in \eqref{costadm} is decomposed as: 

\begin{equation}\label{augcost}
\min f(\mathbf{x})+g(\mathbf{\omega}) \quad \text{ subject to } \quad \mathbf{x}-\mathbf{B}\mathbf{\omega}=0
\end{equation}

\noindent where 
\begin{equation}
f(\mathbf{x})=\lambda_{2}\Vert\mathbf{x}\Vert_{1}
\end{equation}
\noindent and

\begin{equation}
g(\mathbf{\omega})=\Vert \mathbf{y}-\mathbf{A}\mathbf{\omega}\Vert_{2}^{2}+\lambda_{1}\Vert \mathbf{\Gamma}\mathbf{\omega}\Vert_{2}^{2}
\end{equation}

The augmented Lagrangian function of the cost in \eqref{augcost} is obtained as:
\begin{equation}
L_{\rho}(\mathbf{x}, \mathbf{\omega},\mathbf{\lambda})=f(\mathbf{x})+g(\mathbf{\omega})+\mathbf{\Lambda}^{T}(\mathbf{x}-\mathbf{B}\mathbf{\omega})+(\rho/2)\Vert \mathbf{x}-\mathbf{B}\mathbf{\omega}\Vert_{2}^{2} 
\end{equation}

The iterative scheme used in ADMM for minimization consists of three updating steps: 
$\mathbf{x}$-minimization step, $\mathbf{\omega}$-minimization step, and $\mathbf{\Lambda}$-updation step.
The $\mathbf{x}$-updation step is :
\begin{equation}
\mathbf{x}^{k+1}=argmin_{\mathbf{x}}L_{\rho}(\mathbf{x},\mathbf{\omega}^{k},\mathbf{\Lambda}^{k})
\end{equation}

\begin{equation}
\mathbf{x}^{k+1}=argmin_{\mathbf{x}}\Vert\mathbf{x}\Vert_{1}+\mathbf{x}^{T}\left(\dfrac{\mathbf{\Lambda}^{k}-\rho\mathbf{B}\mathbf{\omega}^{k}}{\lambda_{2}}\right)+\dfrac{\rho}{2\lambda_{2}}\Vert\mathbf{x}\Vert_{2}^{2} 
\end{equation}

Solving the above problem, we get the $\mathbf{x}$-updation step as:
\begin{equation}
\mathbf{x}^{k+1}=\dfrac{\lambda_{2}}{\rho}S_{\alpha}\left(\dfrac{\rho\mathbf{B}\mathbf{\omega}^{k}-\mathbf{\Lambda}^{k} }{\lambda_{2}}\right)
\end{equation}

The second step is $\omega$-updation step as:
\begin{equation}
\mathbf{\omega}^{k+1}=argmin_{\mathbf{\omega}}L_{\rho}\left(\mathbf{x}^{k+1},\mathbf{\omega},\mathbf{\Lambda}^{k}\right) 
\end{equation}

Thus, we get:
\begin{equation}
\begin{array}{c}
\mathbf{\omega}^{k+1}=argmin_{\mathbf{\omega}}\Vert\mathbf{y}-\mathbf{A}\mathbf{\omega}\Vert_{2}^{2}+\lambda_{1}\Vert\mathbf{\Gamma}\mathbf{\omega}\Vert_{2}^{2}-\mathbf{\Lambda}^{T}\mathbf{B}\mathbf{\omega}+\\
\dfrac{\rho}{2}\Vert\mathbf{B}\mathbf{\omega}\Vert_{2}^{2}-\rho\mathbf{x}^{T}\mathbf{B}\mathbf{\omega}
\end{array}
\end{equation}
This is a modified least squares algorithm which is solved as:
\begin{equation}
\begin{array}{c}
\mathbf{C}=\left(2\mathbf{A}^{T}\mathbf{A}+2\lambda_{1}\Gamma ^{T}\Gamma+\rho \mathbf{B}^{T}\mathbf{B} \right)\\
\mathbf{\omega}^{k+1}=\mathbf{C}^{-1}\left( \left( \mathbf{z}^{K}+\rho\mathbf{x}^{K+1}\right)^{T}\mathbf{B}+2\mathbf{A}^{T}\mathbf{y} \right) 
\end{array}
\end{equation}

The $\mathbf{\Lambda}$-updation step is as:
\begin{equation}
\mathbf{\Lambda}^{k+1}=\mathbf{\Lambda}^{k}+\rho \left(\mathbf{x}^{k+1}-\mathbf{B}\mathbf{\omega}^{k+1}\right) 
\end{equation}

The reconstruction algorithm can be summarized as illustrated in Algorithm \ref{mh-admm ALG}.

\begin{algorithm}
\caption{Multi-Hypothesis algorithm using Sparsity and Tikhonov regularization (MH-ST)}\label{mh-admm ALG}
\begin{algorithmic}[1]
\State \textbf{input:}
\State{A measurement matrix } $\mathbf{\Phi} \in \mathbb{R}^{m\times n}$
\State{A measurement vector } $\mathbf{y}\in \mathbb{R}^{m}$
\State{maximum number of iterations } $K_{max}$
\State \textbf{output:}
\State{A recovered estimate }$\widehat{\mathbf{x}}\in \mathbb{R}^{n}${ of the original signal.}
\Procedure{MH-ST}{${\mathbf y},{\mathbf x}$}
\State$ \mathbf{x}^{(1)}\gets 0$
\State$\mathbf{z}^{(1)}\gets 0$
\State$\mathbf{\omega}^{(1)}\gets 0$
\State$\mathbf{C}\gets\left(2\mathbf{A}^{T}\mathbf{A}+2\lambda_{1}\Gamma ^{T}\Gamma+\rho \mathbf{B}^{T}\mathbf{B} \right)^{-1}$
\For{$K=1\ldots K_{max}$}
\State${\mathbf{x}}^{(K+1)}\gets \dfrac{\lambda_{2}}{\rho} S_{1}\left(\dfrac{\rho \mathbf{B}\mathbf{\omega}^{K}-\mathbf{z}^{K}}{\lambda_{2}}\right)$
\State$\mathbf{\omega}^{K+1}\gets\mathbf{C}\left( \left( \mathbf{z}^{K}+\rho\mathbf{x}^{K+1}\right)^{T}\mathbf{B}+2\mathbf{A}^{T}\mathbf{y} \right) $
\State$\mathbf{z}^{K+1}\gets \mathbf{z}^{K}+\rho\left( \mathbf{x}^{K+1}-\mathbf{B}\mathbf{\omega}^{K+1}\right) $
\EndFor
\State$\tilde{\mathbf{\omega}}\gets \mathbf{z}^{K+1}$
\State$\tilde{\mathbf{x}}\gets\mathbf{H}\tilde{\mathbf{\omega}}$
\State \textbf{return} $\quad \tilde{\mathbf{x}}$
\EndProcedure
\end{algorithmic}
\end{algorithm}

%
%

where $S_{\alpha}$ is the shrienkage function defined as:
\begin{equation}
S_{\alpha}(\mathbf{x})=\left\lbrace 
\begin{array}{c}
\mathbf{x}+\alpha \quad \mathbf{x}<-\alpha\\

\mathbf{x}-\alpha \quad \mathbf{x}>\alpha
\end{array}\right. 
\end{equation}

As the cost function in \eqref{op1} is convex, the ADMM technique is proved to converge to its global minimum \cite{han2012note, boyd2011distributed}.

In order to enhance the recovered image, we adopt the residual recovery technique as applied in \cite{chen2013elastic}. The recovered signal is sampled and subtracted from the measurement vector to produce a measurement of the residual vector which is assumed to be sparse in the signal domain due to the high correlation of the recovered signal with the original one. The measurment of the residual vector is obtained as:
\begin{equation}
\begin{aligned}
\mathbf{y}_{r}=&\mathbf{y}-\mathbf{\Phi}\tilde{\mathbf{x}}\\
=&\mathbf{\Phi}\left(\mathbf{x}-\tilde{\mathbf{x}}\right) \\
=&\mathbf{\Phi}\mathbf{\Delta x}\\
\end{aligned}
\end{equation}
where $\mathbf{\Delta x}$ and $\mathbf{y}_{r}$ represent for the residual vector, and the measurement of the residual vector, respectively. Assuming that $\mathbf{\Delta x}$ is sparse, we can recover it from its measurements. We apply the BCS-SPL recovery technique in this case to reconstruct the vector $\widehat{\mathbf{\Delta x}}$. The estimated residual vector is then added to the recovered signal of the first stage to produce an enhanced recovery of the signal that is $\widehat{\mathbf{x}}$:
\begin{equation}
\widehat{\mathbf{x}}=\tilde{\mathbf{x}}+\widehat{\mathbf{\Delta x}}
\end{equation}

\section{Simulation Results}
In this section, the simulation results are reported. Five standard video sequences \textit{news}, \textit{foreman}, \textit{football}, and \textit{mother and daughter} are tested. The \textit{foreman} and \textit{football} are fast sequences with abrupt changes. The \textit{news} and \textit{mother and daughter} sequences are slow motion videos with tiny changes. Two frames are considered as the reference which are the first and the last frames of each group. Our proposed algorithm is a compressed video sensing method which works based on the MH motion compensation technique. The two techniques, Multi-Hypothesis Tikhonov (MH-Tikhonov) proposed in \cite{5749477} and Multi-Hypothesis Elasticnet (MH-Enet) proposed in \cite{chen2013elastic} are the CS-based MH methods that are simulated as a benchmark for our proposed scheme. 
We would simulate the mentioned CVS-MH techniques and compare them with the proposed MH-ST algorithm for equal factors such as search window size. 

\subsection{Setting the parameters}
\begin{figure}[h!]
    \centerline{\epsfig{figure=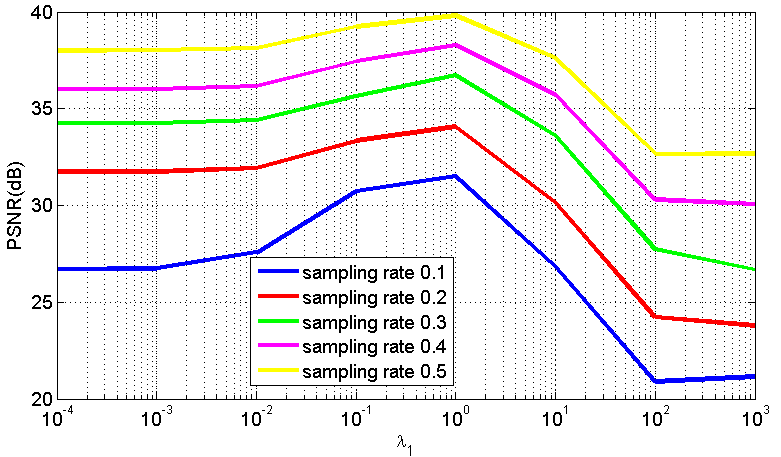,width=9cm}}
    \caption{Reconstruction quality of the $5^{th}$ frame of \textit{foreman} sequence using the MH-ST with different $\lambda_{1}$ at various sampling rates where $\lambda_{2}=1000$, and $\rho=0.01$.}
    \label{fig1}
\end{figure}

In this subseciton, we describe how the parameters of the proposed method are set. We set $K_{max}=10$ since this value provides the convergence of the algorithm. The algorithm is not too sensitive to $\lambda_{2}$, thus we set $\lambda_{2}=1000$. There are two other paramters in the algorithm that should be adjusted: $\lambda_{1}$, and $\rho$. We fix one parameter and run the algorithm for different values of the other one. The resultant Peak Signal to Noise Ratio (PSNR) is observed for various sampling rates and the parameter is adjusted, accordingly. 
It should be noted that the parameter setting tests are done on the \textit{foreman} video sequence. We have also simulated the other sequences and the results are the same. 
Figure \ref{fig1} depicts the PSNR versus $\lambda_{1}$ curve for various sampling rates. The other parameter is fixed as:  $\rho=0.01$. 
Since the parameter $\lambda_{1}$ controls the effect of the Tikhonov regularization term, it is expected to have a small value. The value of $\lambda_{1}$ for which the algorithm achieves its maximum PSNR in all of the sampling rates is selected as the optimum value.

According to the figure, $\lambda_{1}=1$ results in the highest PSNR for all of the sampling rates.


%

\begin{figure}[h!]
    \centerline{\epsfig{figure=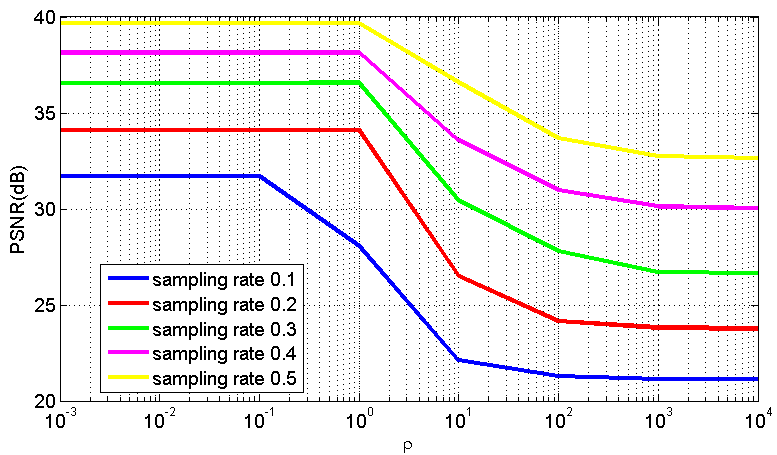,width=9cm}}
    \caption{Reconstruction quality of the $5^{th}$ frame of \textit{foreman} sequence algorithm using the MH-ST with different $\rho$ at various sampling rates where $\lambda_{1}=1$, and $\lambda_{2}=1000$.}
    \label{fig3}
\end{figure}

Figure \ref{fig3}, on the other hand, is the plot of PSNR versus $\rho$ when $\lambda_{1}$ is fixed to 1.

Similar to the previous case, the optimum value for $\rho$ is at the highest PSNR for all of the sampling rates. According to this figure, any value smaller than $0.1$ is proper for $\rho$. However, since the smaller values of $\rho$ bring about a smaller rate of convergence, we set $\rho=0.01$. 

\subsection{The simulations of video sequences}\label{sim2sec}

In this part, various simulations are reported. The proposed method is simulated in this subsection using the parameters adjusted in the previous part. At first, we analyze the algorithm for recovery of a single frame of a group. Then, we would investigate the performances of the algorithms in the case of reconstruction of a large number of frames. To have better intution of the algorithms, we consider two scenarios which use different number of blocks of the search window. The first one is for $p=20$ and the second is for $p=400$.
The measurement matrix used for sampling the blocks of the frames is random Gaussian matrix of size $m \times n$ where $m/n$ indicates the sampling rate. 
In order to achieve better recovery performance of the sequences, the first and the last frames of each group are considered as the reference frames. The reference frame is compressively sampled with a rate of $0.7$ and sent to the decoder where the BCS-SPL recovery technique is adopted for reconstruction. The blocks of the current frame are sampled using the measurement matrix and sent to the decoder. The proposed method together with the other benchmark techniques are simulated to recover the non-reference frames in the receiver side.

As the first set of tests, we exhibit the results for a single frame recovery. For all of the sequences \textit{news}, \textit{football}, \textit{foreman}, and \textit{mother and daughter}, the frames $1$ and $9$ are considered as the reference according to which the $5^{\emph{th}}$ frame is recoverd in the receiver. The corresponding PSNR curves are depicted versus sampling rate for the proposed method together with MH-Tikhonov and MH-Enet methods. 

\begin{figure}[h!]
    \centerline{\epsfig{figure=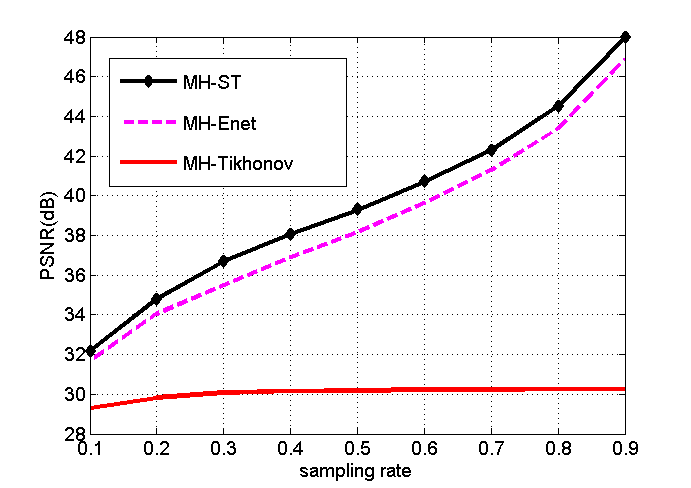,width=9cm}}
    \caption{Recovery performance for the $5^{\emph{th}}$ frame of the \textit{foreman} sequence for $p=20$}
    \label{fig5}
\end{figure}

Figure \ref{fig5} depicts the PSNR versus sampling rate curves for the \textit{foreman} sequence using various techniques.

\begin{figure}[h!]
    \centerline{\epsfig{figure=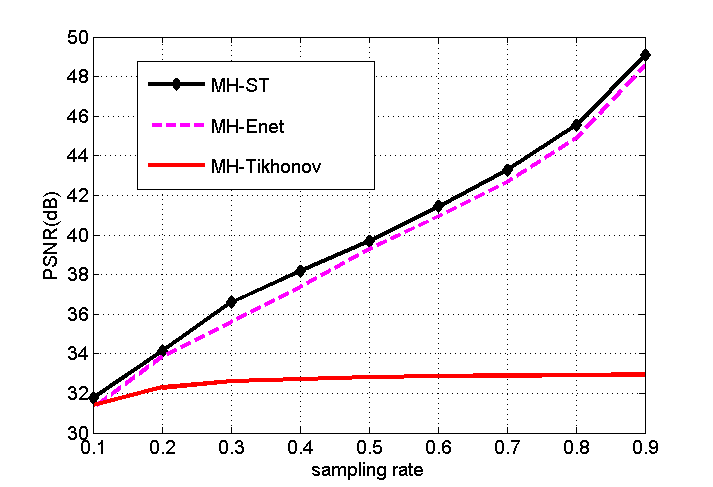,width=9cm}}
    \caption{Recovery performance for the $5^{\emph{th}}$ frame of the  \textit{news} sequence for $p=20$}
    \label{fig6}
\end{figure}

\begin{figure}[h!]
    \centerline{\epsfig{figure=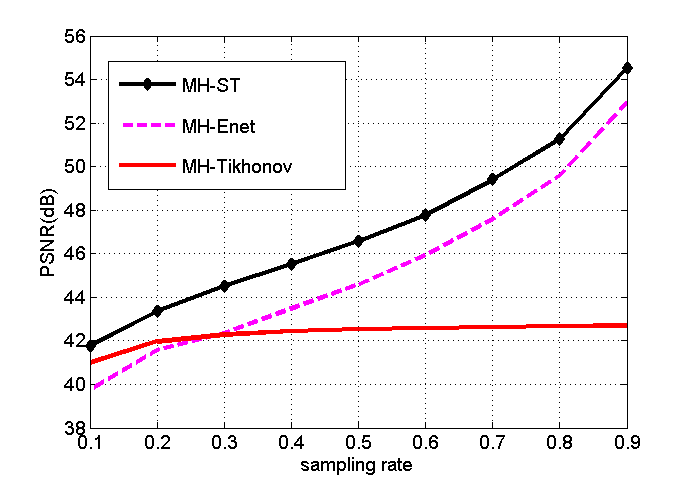,width=9cm}}
    \caption{Recovery performance for the $5^{\emph{th}}$ frame of the \textit{mother and daughter} sequence for $p=20$}
    \label{fig7}
\end{figure}

 It is obvious from the figure that the proposed method considerably outperforms the other two techniques. The PSNR of the MH-ST method is on average $1$ dB and $9.55$ dB more than those of MH-Enet and  MH-Tikhonov, respectively. The similar PSNR-sampling rate curves are exhibited in Figures \ref{fig6} and \ref{fig7} for the \textit{news} and \textit{mother and daughter} sequences, correspondingly.

\begin{figure}[h!]
    \centerline{\epsfig{figure=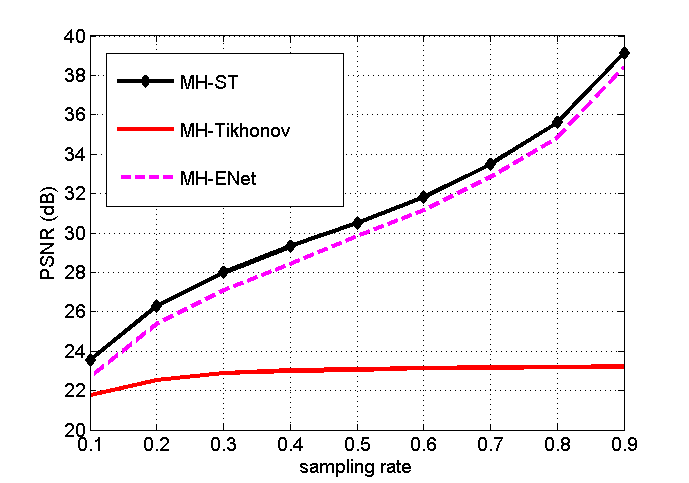,width=9cm}}
    \caption{Recovery performance for the $5^{\emph{th}}$ frame of the \textit{football} sequence for $p=20$}
    \label{fig8}
\end{figure}

The results of Figures \ref{fig6} and \ref{fig7} indicate that the proposed MH-ST technique is better than the MH-Tikhonov and MH-Enet schemes. For the \textit{news} sequence, the PSNR of the proposed method is $0.58$ dB more than that of the MH-Enet, and $7.35$ dB better than the result of MH-Tikhonov, averaged over all the sampling rates. Also for the \textit{mother and daughter} sequence, the MH-ST method achieves $4.87$ dB improvement in PSNR compared to MH-Tikhonov and $1.87$ dB improvement compared to MH-Enet. 
As an other challenging example, we test the fast sequence of \textit{football}. The corresponding PSNR curve is depicted in Figure \ref{fig8}.

According to this figure, the proposed method surpasses the other two techniques, \emph{i.e.}, it has $0.77$ dB improvement over the MH-Enet and $7.95$ dB improvement over the MH-Tikhonov. In the above results, we observed that the MH-ST method outperforms the MH-Enet scheme, and has significant improvement compared to MH-Tikhonov. 
The overall conclusion of the simulation results of this scenario $(p=20)$ is as follows: The MH-Tikhonov technique is not so much efficient in this case. The MH-ST method performs better than the MH-Enet technique. The outperformance of the proposed MH-ST technique over the other two schemes becomes more evident in the case of fast video sequences which are more sophisticated. The recovery of fast video sequences such as \textit{football} and \textit{foreman} becomes more elaborated since the blocks move rapidly and it gets harder to estimate the current block based on the blocks of the reference frame.

\begin{figure}[h!]
    \centerline{\epsfig{figure=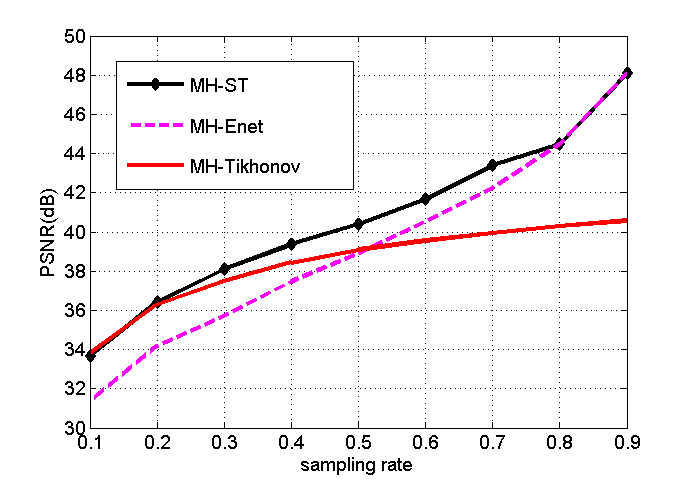,width=9cm}}
    \caption{Recovery performance for the $5^{\emph{th}}$ frame of the \textit{foreman} sequence for $p=400$}
    \label{fig9}
\end{figure}

\begin{figure}[h!]
    \centerline{\epsfig{figure=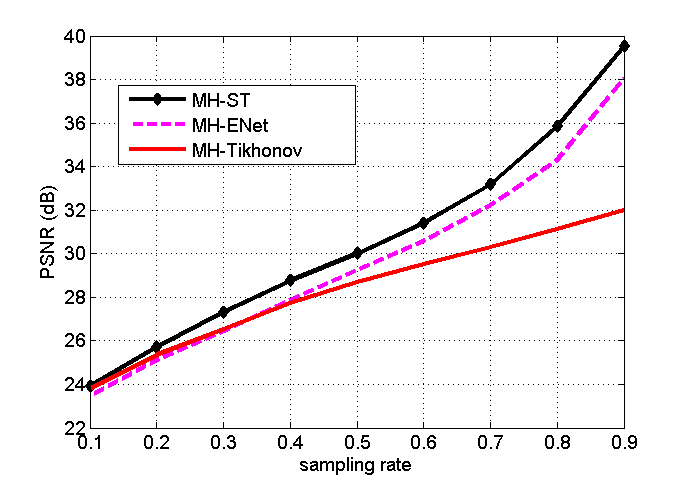,width=9cm}}
    \caption{Recovery performance for the $5^{\emph{th}}$ frame of the \textit{football} sequence for $p=400$}
    \label{fig10}
\end{figure}

In the second scenario, a larger set of reference blocks of the search window is considered so that $p=400$. Since the number of reference blocks estimating the current block is increased in this case, we expect to see better recovery perfromance of the algorithms. Moreover, the increase in the dimension of the vectors makes the problem more complex and the comparison of the computational complexities of the algorithms become vital in this scenario. The PSNR versus sampling rate curves of the \textit{foreman} and \textit{football} sequences are depicted in Figures \ref{fig9} and \ref{fig10}.

 According to these figures, the proposed MH-ST technique behaves better than MH-Enet and MH-Tikhonov. For the \textit{foreman} sequence, the PSNR improvement of MH-ST over MH-Enet and MH-Tikhonov is on-average $1.38$ dB and $2.23$ dB, respectively. In the case of the \textit{football} sequence, the proposed method attains PSNR rise of $0.92$ dB and $2.29$ dB over MH-Enet and MH-Tikhonov, correspondingly. Another fact that can be observed from these two figures is that the MH-Tikhonov offers similar recovery as ours in the low sampling rate of $0.1$, while its performance become inferior to ours as the sampling rate increases.

\begin{figure}[h!]
    \centerline{\epsfig{figure=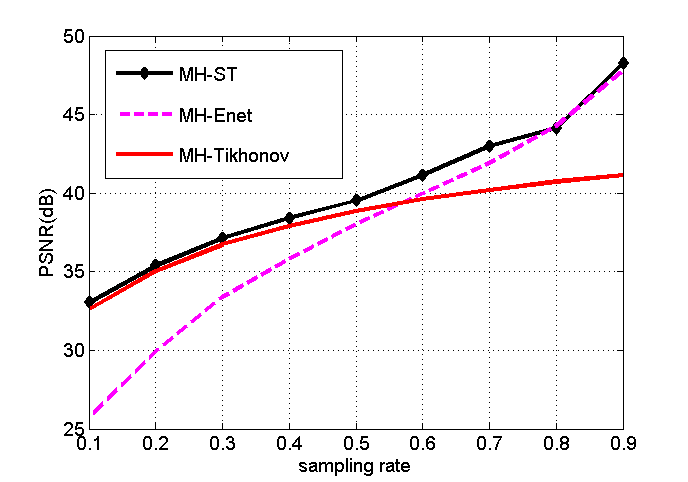,width=9cm}}
    \caption{Recovery performance for the $5^{\emph{th}}$ frame of the \textit{news} sequence for $p=400$}
    \label{fig11}
\end{figure}

\begin{figure}[h!]
    \centerline{\epsfig{figure=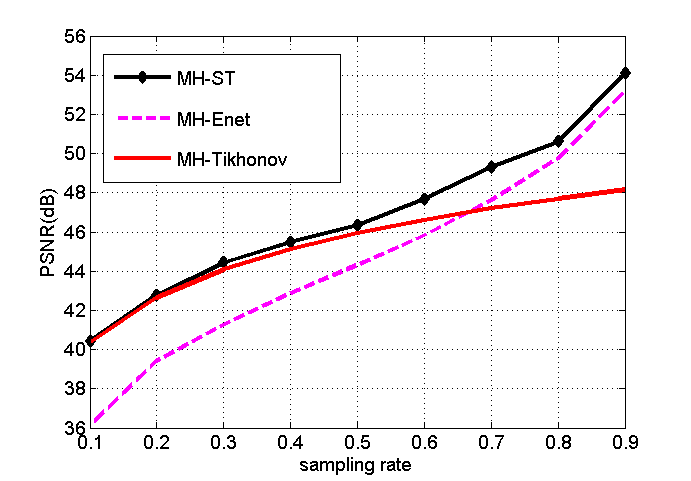,width=9cm}}
    \caption{Recovery performance for the $5^{\emph{th}}$ frame of the \textit{mother and daughter} sequence for $p=400$}
    \label{fig12}
\end{figure}

The outperformance of the MH-ST technique over the other two for the \textit{news} and \textit{mother and daughter} sequences are exhibited in Figures \ref{fig11} and \ref{fig12}.

As can be seen from Figure \ref{fig11}, the PSNR of MH-ST is on-average $2.57$ dB and $1.9$ dB more than those of MH-Enet and MH-Tikhonov, while for the \textit{mother and daughter} sequence, these values are $2.29$ dB and $1.47$ dB, respectively. The simulations of this scenario indicated that the performances of all the three techniques have been improved compared to the previous scenario. It is because more number of reference blocks are considered in the matrix $H$. The proposed technique surpasses the other two techniques in the quality of the recovered frame. In this scenario, for very low sampling rates, the MH-Tikhonov behaves in a similar manner as MH-ST, while its performance degrades as increasing the sampling rate. Additionally, the outperformance of the suggested scheme becomes more noticable in the reconstruction of the faster sequences. 

The complexities of the algorithms are another important issue that should be carefully investigated. To this goal, we depict the corresponding CPU time of the 4 simulations above as a measure of complexity in Table \ref{table:w400tim}.

\begin{table}[ht]
\caption{CPU time for $p=400$} 
\centering 
\begin{tabular}{lllllll} 
\hline\hline 
sequence & method&\multicolumn{5}{c}{sampling rate}\\
               &              &  \\
               &             &0.1 & 0.3& 0.5 & 0.7& 0.9
\\ [0.5ex] 
\hline 
coastguard &MH-Tikhonov&5.85& 6.36& 6.75&7.02&7.64\\
                        & MH-Enet & 55.56  & 58.94  & 63.27   & 65.69  & 67.97 \\
                                 & MH-ST & 7.04 & 7.42 &  8.22   & 8.6 &   8.23 \\
foreman& MH-Tikhonov & 7.41  & 6.97 &   7.30 &  8.04 & 8.16\\
                                 & MH-Enet & 58.75 & 62.00 & 65.03 &  67.90 & 70.09 \\
                                 & MH-ST & 9.13  & 7.58 & 8.61& 8.50  & 8.42 \\
mother& MH-Tikhonov & 5.79 & 6.48  & 6.60  & 7.26 & 7.59\\
and daughter                            & MH-Enet & 56.96 & 59.68 & 62.49  & 66.40   & 68.10 \\
                                 & MH-ST & 11.22  & 11.40 & 11.84 & 11.72  & 12.82 \\
news& MH-Tikhonov & 8.74 & 8.75 & 10.06 & 10.06& 10.25\\
                                & MH-Enet & 60.45& 63.93 & 67.62 & 70.72 & 73.17 \\
                                 & MH-ST & 9.49 & 9.01& 9.3 & 9.2& 9.45\\
football& MH-Tikhonov & 4.90 & 5.54 & 5.19 & 6.01 & 5.76\\
                                & MH-Enet & 56.05 & 56.89 & 59.80 & 61.47 & 62.51 \\
                                 & MH-ST & 8.30 & 8.12 & 9.22 & 8.44 & 9.45 \\ [1ex] 
\hline 
\end{tabular}
\label{table:w400tim} 
\end{table}

It is obvious from Table \ref{table:w400tim} that the CPU time of the proposed technique is slightly more than that of the simplest technique, MH-Tikhonov. The MH-Enet scheme, however, is much more complex than the proposed method. Its CPU time is around $7$ times that of the MH-ST technique. Hence, compared to its other counterparts, the suggested scheme is computationally efficient besides exhibiting better recovery performance. 

In another test, we adopt the three aforementioned techniques to recover a large number of frames (4 groups of 9 frames). The average PSNR of the 36 frames is computed for all the methods and different video sequences and the results are depicted in Table \ref{table:AVGPSNRw20}.
\begin{table}[ht]
\caption{Average PSNR for $p=20$} 
\centering 
\begin{tabular}{lllllll} 
\hline\hline 
sequence & method & \multicolumn{5}{c}{sampling rate}
\\
       &              &  \\
               &             & 0.1 & 0.3 &0.5 &0.7& 0.9 
\\ [0.5ex] 
\hline 
coastguard               & MH-Tikhonov & 24.42 & 25.69 & 25.97 & 26.07 & 26.14\\
                                 & MH-Enet & 24.79 & 28.13 & 30.94 & 34.05 & 39.97 \\
                                 & MH-ST & 25.54 & 28.62 & 31.17 & 34.27 & 40.35 \\
foreman                    & MH-Tikhonov & 28.97 & 29.93  & 30.08  & 30.15 & 30.18\\
                                 & MH-Enet & 30.06 & 34.08 & 37.03 & 40.22 & 46.16 \\
                                 & MH-ST & 30.64 & 35.20 & 37.81 & 40.83  & 46.83 \\
mother & MH-Tikhonov & 40.70 & 42.00 & 42.28 & 42.38 & 42.43\\
and daughter                                & MH-Enet & 39.56 & 42.11 & 44.57 & 47.57 & 52.94 \\
                                 & MH-ST & 41.29 & 44.29 & 46.54  & 49.31 & 54.51 \\
news                         & MH-Tikhonov & 32.14 & 33.23  & 33.49  & 33.58  & 33.63\\
                                & MH-Enet & 32.07 & 35.12 & 39.24 & 42.82 & 48.60 \\
                                 & MH-ST & 32.48 & 36.25 & 39.78 & 43.35 & 49.17 \\
football                     & MH-Tikhonov & 18.26 & 18.82 & 18.92 & 18.97 & 18.98\\
                                & MH-Enet & 21.74 & 26.25  & 29.25 & 32.41 & 38.07 \\
                                 & MH-ST & 22.48 & 27.13 & 29.86 & 33.04 & 39.02 \\ [1ex] 
\hline 
\end{tabular}
\label{table:AVGPSNRw20} 
\end{table}

According to this table, for the first scenario of $p=20$, the average PSNR over $36$ frames obtained by MH-ST technique exceeds what achieved by the other schemes. A similar test is conducted regarding the second scenario of $p=400$. The average PSNR of the three methods for various video sequences is represented in Table \ref{table:AVGPSNRw400}.

\begin{table}[ht]
\caption{Average PSNR for $p=400$} 
\centering 
\begin{tabular}{lllllll} 
\hline\hline 
sequence & method & \multicolumn{5}{c}{sampling rate}
\\
                 &           &                                       \\
               &             & 0.1 & 0.3 & 0.5& 0.7& 0.9
\\ [0.5ex] 
\hline 
coastguard               & MH-Tikhonov & 27.44 & 31.21 & 33.78 & 35.86 & 37.93\\
                                 & MH-Enet & 25.76 & 30.18 & 33.49 & 36.85 & 42.75 \\
                                 & MH-ST & 27.43 & 31.36 & 34.13 & 37.22 & 42.35 \\
foreman                    & MH-Tikhonov & 33.80 & 37.35 & 39.10  & 40.15 & 40.92\\
                                 & MH-Enet & 30.87 & 35.28  & 38.34 & 41.59   & 47.44 \\
                                 & MH-ST & 33.58 & 37.73 & 39.91 & 42.89 & 47.81 \\
mother & MH-Tikhonov & 40.32 & 44.13 & 46.12  & 47.44 & 48.44\\
and daughter                                   & MH-Enet & 36.14  & 41.14 & 44.33  & 47.63  & 53.14 \\
                                 & MH-ST & 40.32 & 44.41 & 46.56 & 49.56 & 54.36 \\
news                         & MH-Tikhonov & 33.12 & 37.17  & 39.49  & 40.85 & 41.90\\
                                & MH-Enet & 25.79 & 33.49 & 38.19  & 42.07 & 48.02 \\
                                 & MH-ST & 33.42 & 37.45 & 39.93 & 43.36  & 48.50 \\
football                     & MH-Tikhonov & 23.24 & 26.56  & 28.88 & 30.59  & 32.21\\
                                & MH-Enet & 22.63 & 26.15& 29.21 & 32.42  & 38.28 \\
                                 & MH-ST & 23.42 & 27.13 & 29.27 & 32.18  & 36.95 \\ [1ex] 
\hline 
\end{tabular}
\label{table:AVGPSNRw400} 
\end{table}

As reported by the table, the offered technique presents higher recovery accuracy in comparison with the other benchmark schemes.  

\begin{figure}[h!]
\centering
    \begin{subfigure}[b]{0.5\textwidth}
\centering
	\includegraphics[width=0.5\textwidth]{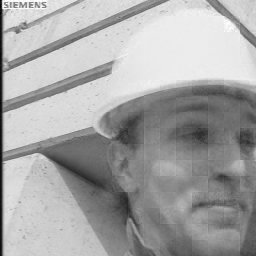}
	\caption{MH-ST}
    \end{subfigure}
    \begin{subfigure}[b]{0.5\textwidth}
\centering
    	\includegraphics[width=0.5\textwidth]{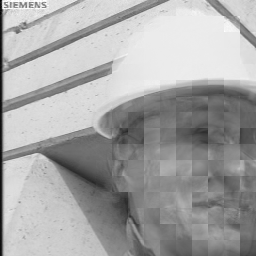}
  	\caption{MH-Tikhonov}
    \end{subfigure}
    \caption{Recovery performance for the $5^{\emph{th}}$ frame of the \textit{foreman} sequence for MH-ST (up) and MH-Tikhonov (down) in the case of $p=20$ and sampling rate of $30\%$.}
\label{test_20}
\end{figure}

In order to have a subjective comparison, the reconstructions of the MH-ST and the MH-Tikhonov techniques for \textit{foreman} sequence are exhibited in Figure \ref{test_20} for sampling rate of $30\%$ and $p=20$.

\begin{figure}[h!]
\centering
    \begin{subfigure}[b]{0.5\textwidth}
	\centering
    \includegraphics[width=0.5\textwidth]{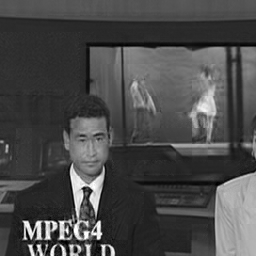}
\caption{MH-ST}
    \end{subfigure}
    \begin{subfigure}[b]{0.5\textwidth}
	\centering
    \includegraphics[width=0.5\textwidth]{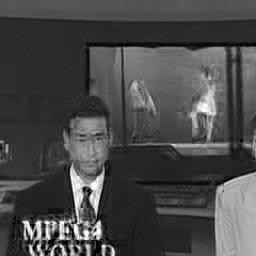}
\caption{MH-Enet}
    \end{subfigure}
    \caption{Recovery performance for the $5^{\emph{th}}$ frame of the \textit{news} sequence for MH-ST (up) and MH-Enet (down) in the case of $p=400$ and sampling rate of $10\%$.}
    \label{test_400}
\end{figure}

To manifest the subjective difference of the MH-ST and MH-Enet schemes, we depict the results of the \textit{news} sequence for the sampling rate of $10\%$ and $p=400$ in Figure \ref{test_400}.

Both of the figures indicate that the proposed MH-ST method has better recovery performance compared to the MH-Enet and MH-Tikhonov. 
These subjective results confirm the objective simulations of PSNR versus rate of Figure 4 and Figure 10. 

\section{Conclusion}
In this paper, a multi-hypothesis compressed video sensing strategy is suggested which exploits the sparsity of the video frames to reconstruct the signal at the decoder side. We defined a cost function consisting of a Tikhonov regularizer and a sparsity promoting term to estimate the non-reference blocks by a linear combination of the reference ones. The extensive simualtions conducted on various video sequences confirm that the proposed MH-ST technique achieves higher reconstruction accuracy of the frames while at the same time being computationally efficient.
\bibliographystyle{ieeetran}
\bibliography{cvs_aexiv}
\end{document}